\documentclass[twoside]{ilcws07}
\usepackage[latin1]{inputenc}
\usepackage[dvips]{graphicx,epsfig,color}
\usepackage{wrapfig,rotating}
\usepackage{amssymb,amsmath,array}

\begin{document}
\title{Probing CP properties of the Higgs Boson via \boldmath{$e^+e^-\to t
\bar{t}\phi$}} %%
%***********************************************************************
% AUTHORS INFORMATION AREA
%***********************************************************************
\author{R.M.~Godbole$^1$, P.S.~Bhupal Dev$^1$, A.~Djouadi$^2$,
M.M.~M\"uhlleitner$^3$ and S.D.~Rindani$^4$
% Optional short acknowledgment: remove next line if non-needed
%\thanks{This is an optional funding source acknowledgment.}
% DO NOT MODIFY THE FOLLOWING '\vspace' ARGUMENT
\vspace{.3cm}\\
% Addresses and institutions (remove "1- " in case of a single institution)
1- Center for High Energy Physics - Indian Institute of Science \\
Bangalore 560 012 - India
%% Remove the next three lines in case of a single institution
\vspace{.1cm}\\
2- Laboratoire de Phyisque Th\'eorique - U.~Paris-Sud and CNRS\\
F-91405 Orsay - France
\vspace{0.1cm}\\
3- Theory Division, Department of Physics - CERN \\
CH-1211 Geneva 23 - Switzerland \\
Laboratoire de Physique Th\'eorique - LAPTH \\
F-74941 Annecy-le-Vieux - France
\vspace{0.1cm}\\
4- Theoretical Physics Division - Physical Research Laboratory, Navrangpura \\
Ahmedabad 380 009 - India
}
%%***********************************************************************
% END OF AUTHORS INFORMATION AREA
%************************************************************************

\maketitle

\begin{abstract}
One of the main endeavors at future high-energy colliders is the search
for the Higgs boson(s) and, once found, the probe of the fundamental
properties. In particular, the charge conjugation and parity (CP) quantum
numbers have to be determined. We show that these are unambiguously
accessible at future $e^+e^-$ colliders through the measurement of the total
cross section and the top polarization in associated Higgs production with
top quark pairs.
\end{abstract}

\section{Introduction}

After the discovery of the Higgs boson - or several Higgs bosons
in extensions beyond the Standard Model (SM) - we have to probe
its properties in order to establish the Higgs mechanism
\cite{Higgs} as responsible for the creation of particle masses
without violating gauge symmetry. In the SM the Higgs mechanism is
implemented by adding one isodoublet complex scalar field which
leads after electroweak symmetry breaking to one single spin zero
CP-even Higgs particle \cite{Higgs,HHG}. In extensions beyond the
SM the Higgs sector can be non-minimal, as {\it e.g.} in the
Minimal Supersymmetric Standard Model (MSSM), which contains five
physical Higgs states, {\it i.e.} two CP-even $h$ and $H$, one
CP-odd $A$ and two charged $H^\pm$ Higgs bosons
\cite{HHG,MSSMbook}. To establish the Higgs mechanism
experimentally we have to determine the Higgs spin, its behavior
under charge and parity transformations, the couplings to gauge
bosons and fermions, and finally the trilinear and quartic Higgs
self-interactions must be measured to reconstruct the Higgs
potential itself. To fulfill this program the Large Hadron
Collider (LHC) analyses \cite{LHC} must be complemented by the
high-precision measurements at a future International
Linear $e^+e^-$ Collider (ILC) \cite{ILC,LHC-ILC,ILCRDR}. \\
We address in this contribution the determination of the Higgs CP
quantum numbers. To do so in an unambiguous way is somewhat
problematic \cite{cpvhiggs}. Observables sensitive to the Higgs
spin-parity such as angular correlations in Higgs decays into
$V=W,Z$ pairs \cite{Barger,CPdecay} or in Higgs production with or
through these states \cite{Barger,CPprod} only project out the
CP-even component of the $HVV$ coupling, even in the presence of
CP violation. In addition, the purely pseudoscalar $AVV$ coupling
is zero at tree-level and is generated only through tiny loop
corrections. In the Higgs couplings to fermions, however, the
CP-even and CP-odd components can have the same magnitude. Here,
the heaviest fermion discovered so far, the top quark, plays a
special role. The $Htt$ coupling is largest being proportional to
the top quark mass due to the Higgs mechanism. At a future ILC the
Higgs boson can therefore be produced with sufficient rate in
associated production with a $t\bar{t}$ pair, $e^+e^-\to t\bar{t}
H$ \cite{ttHpaper0,ttHpaper}. We propose a simple and
straightforward way to determine the CP nature of a SM-like Higgs
boson in this process, in an unambigious way, where we exploit
that the cross section as well as the top quark polarization
behave in a radically different way for CP-even and CP-odd Higgs
production.
\section{The total production cross section}
\begin{wrapfigure}{r}{0.6\columnwidth}
%\begin{figure}[h]
\centerline{\includegraphics[width=0.67\columnwidth,bb=3 680 600 735]{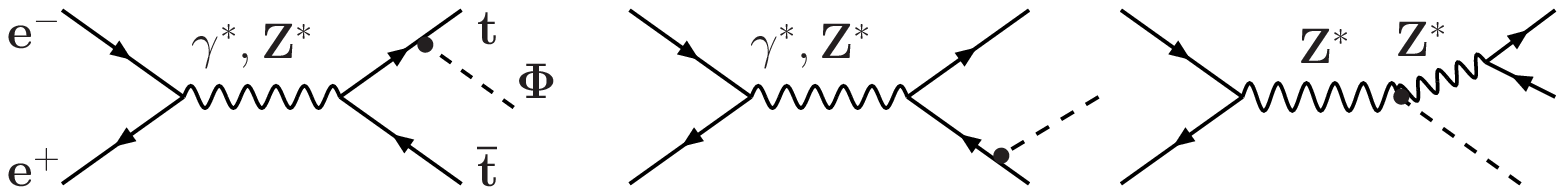}}
\caption{Feynman diagrams for the associated production of Higgs bosons with
a top quark pair.}\label{Fig:diagrams}
\end{wrapfigure}

\begin{wrapfigure}{r}{0.5\columnwidth}
%\begin{figure}[!h]
\begin{center}
\includegraphics[width=0.55\columnwidth,bb=73 485 680 700]{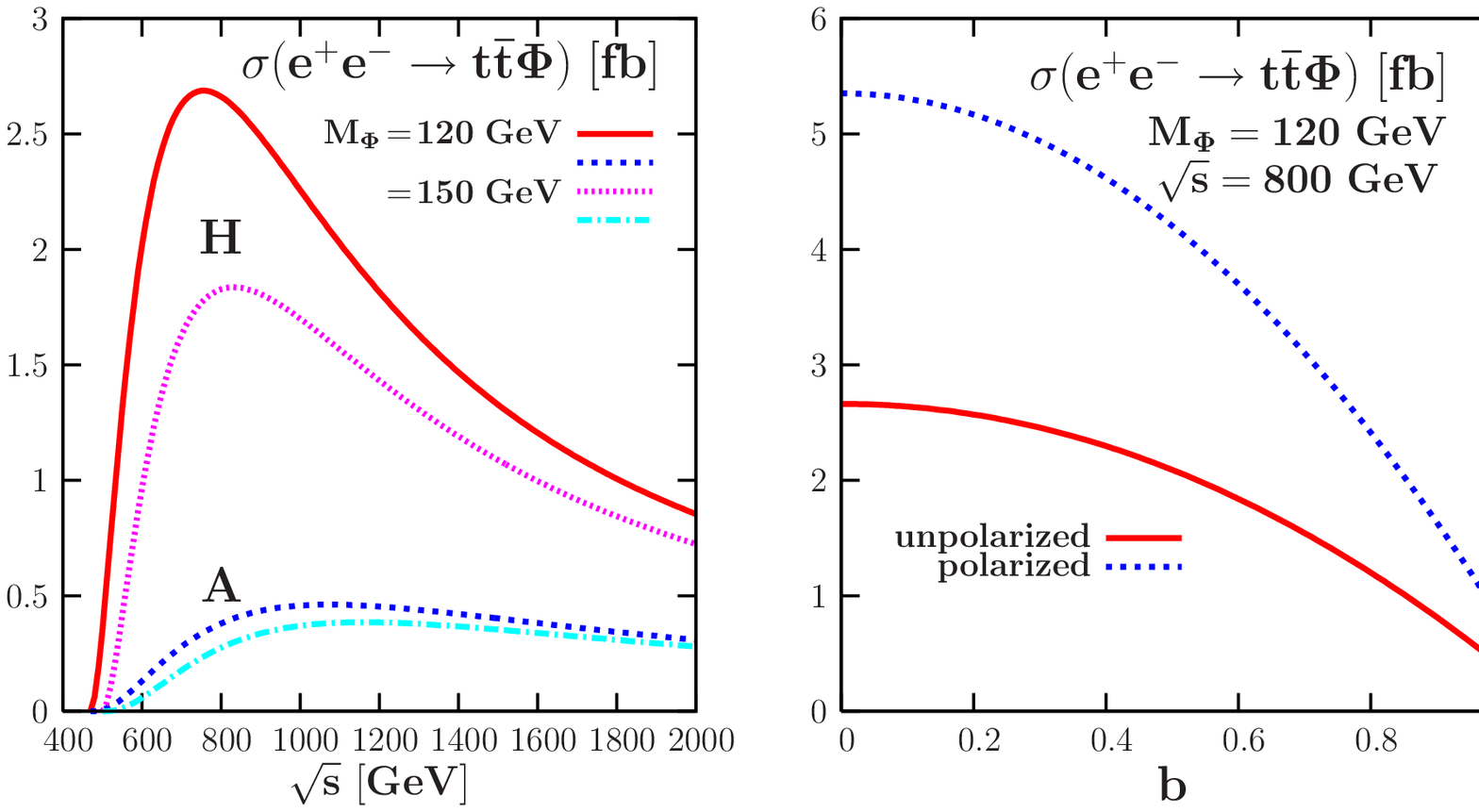}
\end{center}
\caption[]{The production cross sections $\sigma(e^+e^- \to t\bar
t \Phi)$ for a scalar and a pseudoscalar Higgs boson as a function
of $\sqrt s$ for two masses $M_\Phi=120$ and 150 GeV (left) and
for unpolarized and polarized $e^\pm$ beams as a function of the
parameter $b$ at $\sqrt s=800$ GeV with $M_\Phi=120$ GeV (right).}
\label{Fig:totalcxn}
%\end{figure}
\end{wrapfigure}

The diagrams which contribute in the SM to the process $e^+e^-\to
t\bar{t} H$ are shown in Fig.\ref{Fig:diagrams}. The bulk of the
cross section is generated when the Higgs is
%\end{figure}
radiated off the heavy top quarks
\cite{ttHpaper}, whereas the Higgs produced in association with a $Z$ boson
which then splits into a $t\bar{t}$ pair provides only a very small
contribution, amounting to a few percent for $\sqrt{s} \le 1$~TeV
%\footnote{The additional diagram in a 2 Higgs doublet model (2HDM) where the $t\bar t$ pair
%originates from the splitting  of a CP--even (odd) scalar particle for
%(pseudo)scalar Higgs production, contributes very little unless $\Phi\to  t\bar
%t$ decays  are allowed.}.
Detailed simulations  have shown the cross section to be measurable
with an accuracy of order 10\% \cite{ttHexperiment}. We will discuss the
case of a SM-like mixed CP Higgs state $\Phi$ and use the general form of
the $t\bar{t}\Phi$ coupling
\begin{eqnarray}
g_{\Phi tt} = -i \frac{e} {s_W} \frac{m_t}{2M_W}  (a +i b\gamma_5) \;,
\end{eqnarray}
where the coefficients $a$ and $b$ are assumed to be real and $s_W \equiv
\sin\theta_W = \sqrt{1- c_W^2}$. In the SM we have $a=1, b=0$ and for a
purely pseudoscalar Higgs boson $a=0, b\neq 0$.  In the pseudoscalar case we
take $b=1$, consistent with a convenient normalization $a^2 + b^2 =1$ chosen
for the general case of a Higgs boson with an indefinite CP quantum number. A
non--zero value for the product $ab$ will hence signal CP violation in the
Higgs sector. For the $ZZ\Phi$ coupling, we will use the form,
\begin{eqnarray}
g_{ZZ\Phi}^{\mu \nu} = -i c (e M_Z/ s_Wc_W) g^{\mu \nu}\;.
\end{eqnarray}
And for the  numerical analysis we chose $c=a$ \cite{eettH} as
$c=1 (0)$ in the case of  a CP-even (odd) Higgs boson. We will
thus have only one free parameter $b$.  However, this simple
parametrization for a SM--like Higgs need not  be true in {\it
e.g.} a general 2HDM, where $a,b$ and $c$ are three independent
parameters. We have calculated the cross section for the
production of a mixed CP Higgs state including the polarization
dependence of the final state top quarks. The lengthy result has
been checked to agree with Ref.~\cite{ttHpaper} for the
unpolarized total cross section. In Fig.~\ref{Fig:totalcxn} left
panel we show the production cross section for a purely scalar
($H$ with $b=0$) and a pseudoscalar ($A$ with $b=1$) Higgs as a
function of the c.m. energy and for two mass values $M_\Phi =120$
and $150$~GeV. As can be inferred from the figure, the threshold
rise of the cross section in the scalar and the pseudoscalar case
is very different. Furthermore, for the same strength of the $\Phi
tt$ couplings, there is an order of magnitude difference between
the $H$ and $A$ cross sections at moderate energies. Only at very
high energies, $\sqrt{s}\gg 1$~TeV, the chiral limit is reached
and the two cross sections become equal,

\noindent up to the small contribution due to the diagram
including the $ZZ\Phi$ coupling. These two features hence provide
an extremely powerful tool to discriminate the CP properties of
the spin zero particle produced together with the top quark pair.
The difference in the threshold behavior of the $A$ and $H$ case
is strong enough so that the cross section measurement at only two
different c.m. energies allows a clear determination of the CP
properties of the $\Phi$ state. {\it e.g.} for $M_\Phi=120$~GeV,
the ratio of the cross sections at $\sqrt{s}=800$ and 500~GeV is
$\sim 63$ and $\sim 7.5$ for the scalar and pseudoscalar case,
respectively. Finally, taking the ratio makes the conclusion
robust with respect to the effect of the top Yukawa coupling, the
higher order radiative corrections \cite{radcor} or systematic
errors in the measurement.

We also studied the $b$ dependence of the total $t\bar{t}\Phi$
production process at a given energy and for fixed $M_\Phi$. Being
a CP even quantity it only depends on $b^2$.
Fig.~\ref{Fig:totalcxn} (right) shows the result for unpolarized
and polarized $e^\pm$ beams. For the latter, we used the standard
ILC values $P_{e^-}=-0.8$ and $P_{e^+}=0.6$, which double the
total rate.

\section{Top quark polarization as a probe of the CP nature of $\Phi$}
\begin{wrapfigure}{r}{0.5\columnwidth}
\begin{center}
\includegraphics[width=0.55\columnwidth,bb=73 485 680 740]{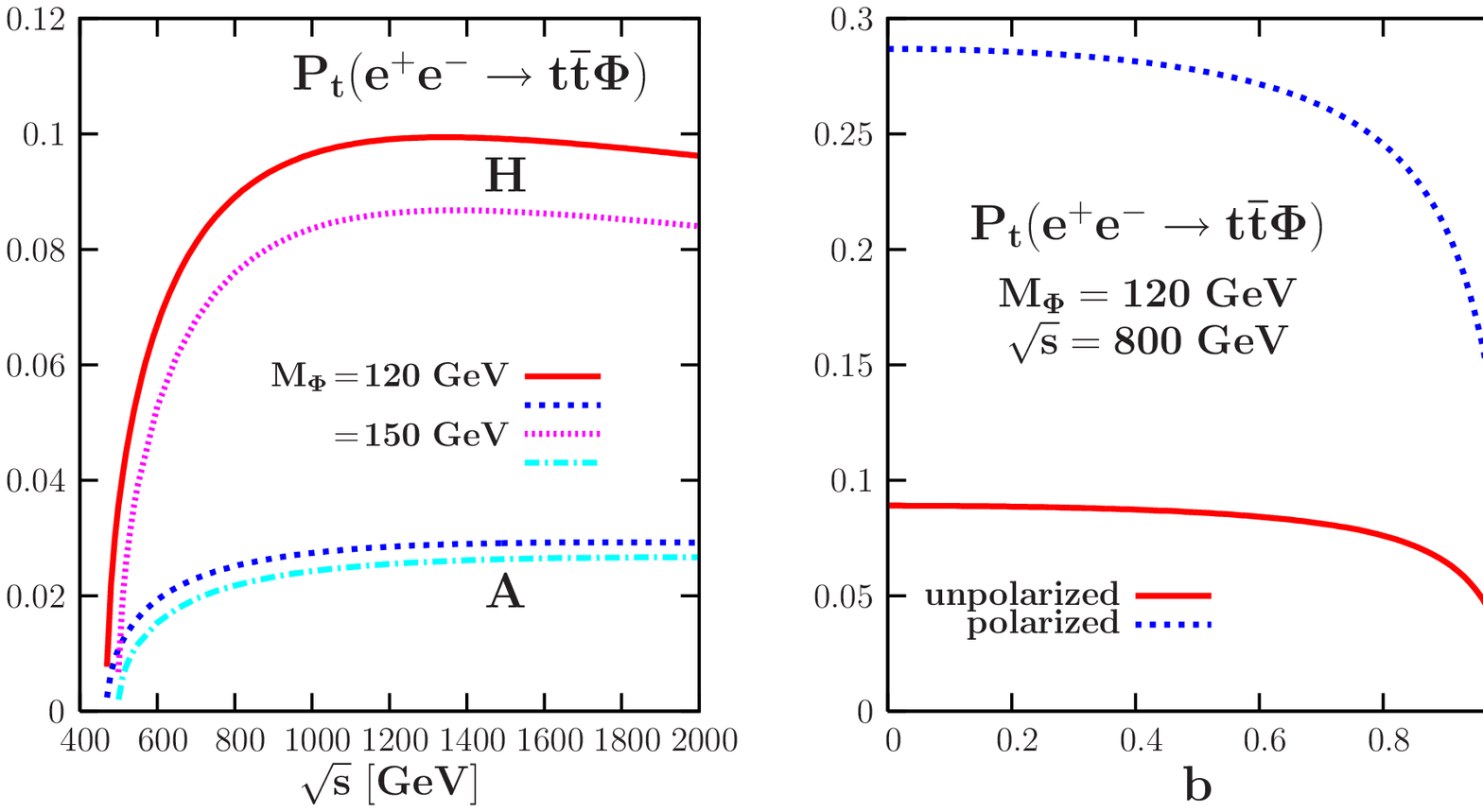}
\end{center}
\caption[]{The top polarization in $e^+e^- \to t\bar t \Phi$
 for a scalar and a pseudoscalar Higgs as a function of $\sqrt s$ for
 $M_\Phi=120$, 150 GeV  (left) and with  unpolarized and
polarized $e^\pm$ beams as a function of $b$ at $\sqrt s=800$ GeV
for $M_\Phi=120$ GeV (right).} \label{Fig:toppol}
\end{wrapfigure}

Since the top quark, due to its large decay width $\Gamma_t\sim
1.5$~GeV, decays much before hadronization, its spin information
is translated to the decay distribution before contamination
through strong interaction effects. Furthermore, the lepton
angular distribution of the decay $t \to bW \to b\bar{l}\nu$ is
independent of any non-standard effects in the decay vertex, so
that it is a pure probe of the physics of the top quark production
process \cite{Tpol}. The net polarization of the top quark
therefore provides an interesting tool for the probe of $b$, see
also Ref.\cite{fermipol}. In Fig.~\ref{Fig:toppol} (left) we show
as a function of $\sqrt{s}$ for $M_\Phi=120$ and $150$~GeV in the
$H(b=0)$ and $A(b=1)$ case the expected degree of $t$-quark
polarization $P_t$, given by
\begin{eqnarray}
P_t = \frac{\sigma(t_L)-\sigma(t_R)}{\sigma (t_L)+\sigma (t_R)} \;.
\end{eqnarray}
As can be inferred from the figure, the degree of top polarization is again
strikingly different for the CP even and CP odd case and shows a very different
threshold dependence.

In addition, since $P_t$ is constructed as a ratio of cross
sections, the insights gained from this variable are not affected
by a possibly model dependent normalization of the overall
$tt\Phi$ coupling strength, higher order corrections etc. $P_t$ is
a parity odd quantity and receives contributions from the
interferences between the $\gamma$ and all $Z$ exchange diagrams,
with the one stemming from the diagram involving the $ZZ\Phi$
vertex being small. The parity violation effect for the emission
of a (pseudo)scalar is controlled by the (vector) axial-vector
$Zt\bar{t}$ coupling ($v_t=(2I_t^{3L} -4Q_t s_W^2)/(4s_w c_W)$)
$a_t=2I_t^{3L}/(4s_W c_W)$, where $I_t^{3L}$ denotes the top
isospin and

\noindent $Q_t$ the electric charge. Hence the ratios of the
$P_t$ values away from threshold are expected to be given by the
$a_t/v_t\sim 3$, which indeed is confirmed by both
Figs.~\ref{Fig:toppol} at $\sqrt{s}=800$~GeV.

\section{The sensitivity to CP mixing}
\begin{wrapfigure}{r}{0.5\columnwidth}
\begin{center}
\includegraphics[width=0.55\columnwidth,bb=73 490 680 700]{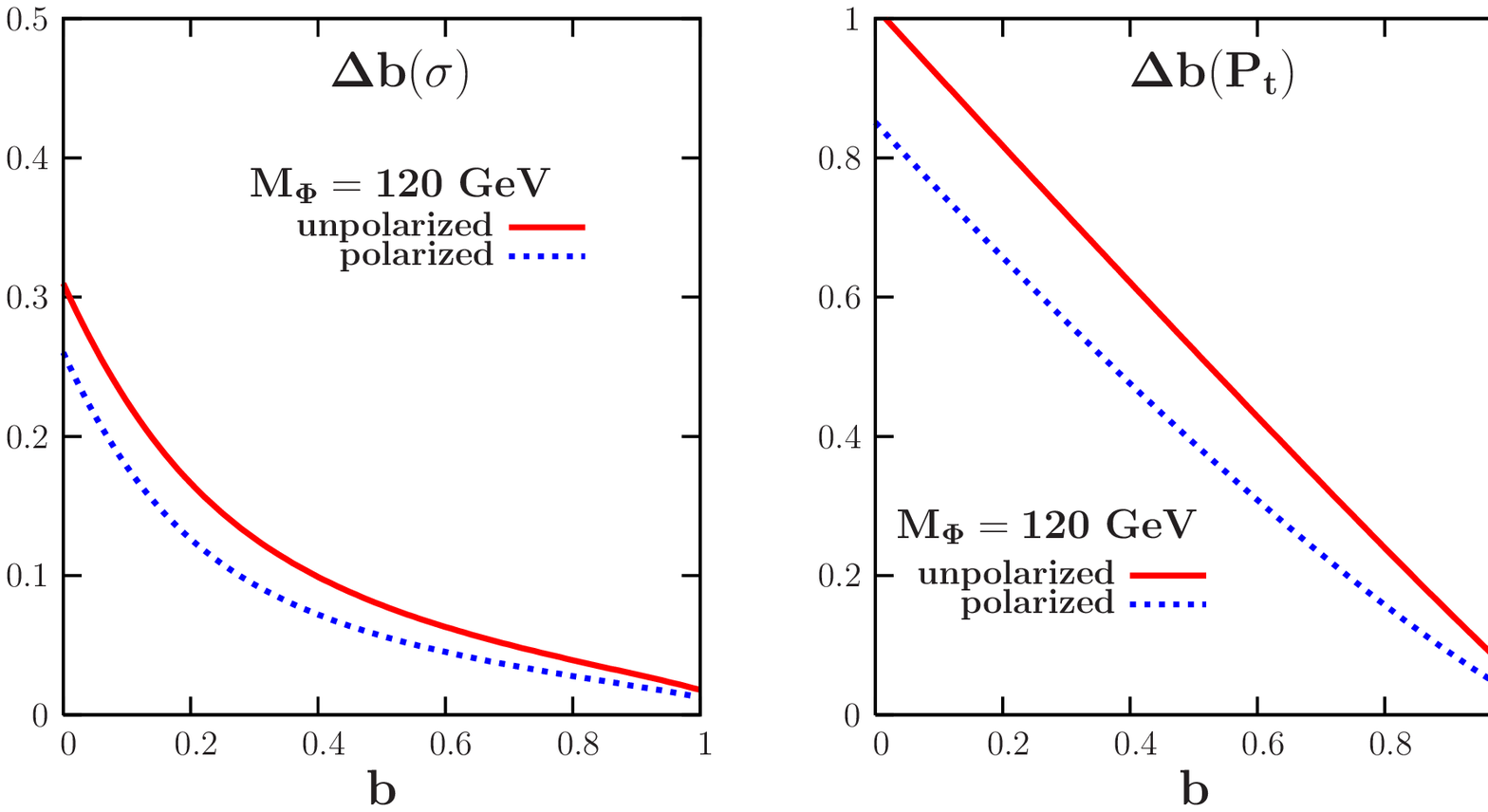}
\end{center}
\caption[]{The sensitivity of the cross section (left) and the top
polarization (right) to $b$ for $M_\Phi=120$ at $\sqrt s=800$
with ${\cal L}=500$ fb$^{-1}$.}
\label{Fig:sens}
\end{wrapfigure}

We investigate how the behavior of the cross section and the
measurement of the top polarization, which both are clear
discriminators between a scalar and pseudoscalar Higgs state, can
be used to get information on the CP mixing, {\it i.e.} the value
of $b$. Ignoring systematical errors, the sensitivity of an
observable $O(b)$ to the parameter $b$ at $b=b_0$ is $\Delta b$,
if $|O(b)-O(b_0)| = \Delta O(b_0)~{\rm for}~|b-b_0| < \Delta b$,
where $\Delta O(b_0)$ is the statistical fluctuation in $O$ at an
integrated luminosity ${\cal {L}}$. For the cross section $\sigma$
and the polarization $P_t$, the statistical fluctuation at a level
of confidence $f$ are given by $\Delta \sigma = f \sqrt{\sigma/
{\cal L}}$ and $ \Delta P_t= f/ \sqrt {\sigma {\cal L} } \times
\sqrt{1-P_t^2}$. Fig.~\ref{Fig:sens} (left) shows the sensitivity
$\Delta b$ from the cross section measurement for

\noindent $M_\phi=120$~GeV at $\sqrt{s}=800$~GeV with ${\cal
L}=500$~fb$^{-1}$. For polarized $e^\pm$ beams it varies from 0.25
for $H(b=0)$ to 0.01 for $A(b=1)$, a rather precise determination
obtained from a simple measurement. The top polarization is less
sensitive to $b$, see Fig.~\ref{Fig:sens} (right).

\begin{wrapfigure}{r}{0.40\columnwidth}
\begin{center}
\vspace*{-1.7cm}
\includegraphics[width=0.40\columnwidth]{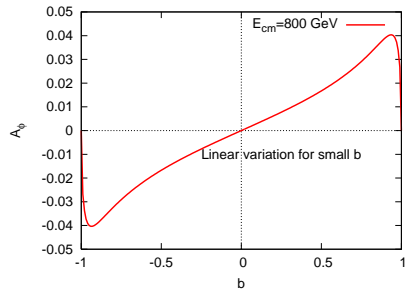}
\end{center}
\vspace*{-0.5cm}
\caption[]{The up-down asymmetry of $\bar{t}$ in associated $t\bar{t}\Phi$
production for $M_\Phi=120$ at $\sqrt s=800$.}
\label{Fig:udasy}
\end{wrapfigure}

Both $\sigma$ and $P_t$ are CP even quantities, they cannot depend
linearly on $b$ and hence not probe CP violation directly.
Observables depending directly on the sine of the azimuthal angle
($\Phi$) are linear in $b$. The up-down asymmetry $A_\Phi$ of the
antitop quark production with respect to the top-electron plane
($\Phi=0$) is an example of such an observable:
% Here the expression has been added
\begin{equation}
A_\Phi=\frac{\sigma_{\rm partial}(0\leq \Phi<\pi)-\sigma_{\rm
partial}(\pi\leq \Phi<2\pi)}{\sigma_{\rm partial}(0\leq
\Phi<\pi)+\sigma_{\rm partial}(\pi\leq \Phi<2\pi)}
\end{equation}
with $\sin\Phi =\frac{(\mathbf p_{e^-}-\mathbf p_{e^+})\cdot
(\mathbf p_t\times \mathbf p'_{\bar t})}{|\mathbf p_{e^-}-\mathbf
p_{e^+}|\cdot|\mathbf p_t\times \mathbf p'_{\bar t}|}$, where
$\mathbf p'_{\bar t}$ is the $\bar t$ momentum in the $\bar
t$-Higgs rest frame. Fig.~\ref{Fig:udasy} shows the asymmetry
$A_\Phi$ for a Higgs boson of 120~GeV and a c.m. energy of
$\sqrt{s}=800$~GeV as a function of $b$. It can reach values of
order 5\%. The non-zero value of the asymmetry arises from the
channel which involves the $ZZ\Phi$ coupling.
%~\cite{barshalom}. 

\section{Summary}
We have shown that the total cross section and the top polarization asymmetry
for associated Higgs production with top quark pairs in $e^+e^-$ collisions
provide a very simple and unambiguous determination of the CP quantum
numbers of a SM-like Higgs particle. Exploiting the up-down asymmetry
of the anti-top with respect to the top-electron plane we further have
a direct probe of CP violation at hand.
\newline

% ****************************************************************************
% BIBLIOGRAPHY AREA
% ****************************************************************************
\begin{footnotesize}
% IF YOU DO NOT USE BIBTEX, USE THE FOLLOWING SAMPLE SCHEME FOR THE REFERENCES
% ----------------------------------------------------------------------------

\end{footnotesize}

% ****************************************************************************
% END OF BIBLIOGRAPHY AREA
% ****************************************************************************

\end{document}